\documentclass[12pt]{iopart}

\usepackage[utf8]{inputenc}
\usepackage[T1]{fontenc}

\usepackage{graphicx}
\usepackage{amsthm}

\theoremstyle{definition}
\newtheorem{example}{Example}
\newtheorem{definition}{Definition}

\newcommand{\ket}[1]{{\left\vert{#1}\right\rangle}}

\begin{document}

\newcommand{\mtt}{Faster manipulation of large quantum circuits using wire label reference diagrams}

\title[Faster manipulation of large quantum circuits using wire label reference diagrams]{Faster manipulation of large quantum circuits\\using wire label reference diagrams}

\author{Alexandru Paler$^{1,2}$, Austin G. Fowler$^3$, Robert Wille$^2$}

\address{$^1$ Linz Institute of Technology, Johannes Kepler University Linz, Austria\\
$^2$ Institute for Integrated Circuits, Johannes Kepler University Linz, Austria\\
$^3$ Google Inc., Santa Barbara, USA
}
\ead{alexandru.paler@jku.at}
\vspace{10pt}
\begin{indented}
\item[]
\end{indented}

\begin{abstract} 
Large scale quantum computing is highly anticipated, and quantum circuit design automation needs to keep up with the transition from small scale to large scale problems. Methods to support fast quantum circuit manipulations (e.g.~gate replacement, wire reordering, etc.) or specific circuit analysis operations have not been considered important and have been often implemented in a naive manner thus far. For example, quantum circuits are usually represented in term of one-dimensional gate lists or as directed acyclic graphs. 
Although implementations for quantum circuit manipulations are often only of polynomial complexity, the sheer number of possibilities to consider with increasing scales of quantum computations make these representations highly inefficient -- constituting a serious bottleneck. At the same time, quantum circuits have structural characteristics, which allow for more specific and faster approaches. This work utilises these characteristics by introducing a dedicated representation for large quantum circuits, namely wire label reference diagrams. We apply the representation to a set of very common circuit transformations, and develop corresponding solutions which achieve orders of magnitude performance improvements for circuits which include up to 80 000 qubits and 200 000 gates. The implementation of the proposed method is available online.
\end{abstract}

\section{Introduction}

Quantum circuits are gradually reaching a point of industrial relevance \cite{mohseni2017commercialize}. Current efforts towards building a large scale fault-tolerant quantum computer are of a hardware nature, where the integration of a large number of qubits is pursued. However, software should stay in line with potential hardware developments. A large scale computer will execute large scale computations expressed as fault-tolerant quantum circuits protected by quantum error correcting codes. Fault-tolerance increases the amount of computational resources required to execute the computation. As a result, the quantum circuits operate on a very high amount of qubits and include large numbers of quantum gates. Consequently, the design automation of corresponding circuits is gaining more and more attention.

In order to ease the design of quantum circuits, various methods for design automation were proposed in the past, including issues of:
\begin{itemize}
	\item synthesis, i.e.~generating initial circuit structures representing the desired quantum functionality (e.g.~\cite{saeedi2013synthesis,shende2003synthesis,DBLP:conf/aspdac/NiemannWD14,miller2003transformation}),
	\item optimisation, i.e.~improving the resulting circuit structure with respect to different (often complementary) objectives such as gate costs (e.g.~\cite{saeedi2013synthesis,DBLP:conf/ismvl/MillerWD10}), number of qubits (e.g.~\cite{DBLP:journals/integration/WilleSMD14,paler2016wire}), or more dedicated objectives such as ensuring nearest neighbour compliance (e.g.~\cite{hirata2011efficient,DBLP:journals/tcad/WilleLD14}), or
	\item decomposition/technology mapping, i.e.~mapping a quantum circuit description e.g.~composed of arbitrary/high level gate libraries to a circuit structure composed of the actually supported elementary operations (e.g.~\cite{barenco1995elementary,NC00,DBLP:conf/rc/NiemannBCJW15}) or a circuit structure following certain schemes, such as the fault-tolerant ICM form (e.g.~\cite{paler2017fault}).
\end{itemize}

\subsection{Motivation}

\begin{figure}
	\includegraphics[width=0.9\columnwidth]{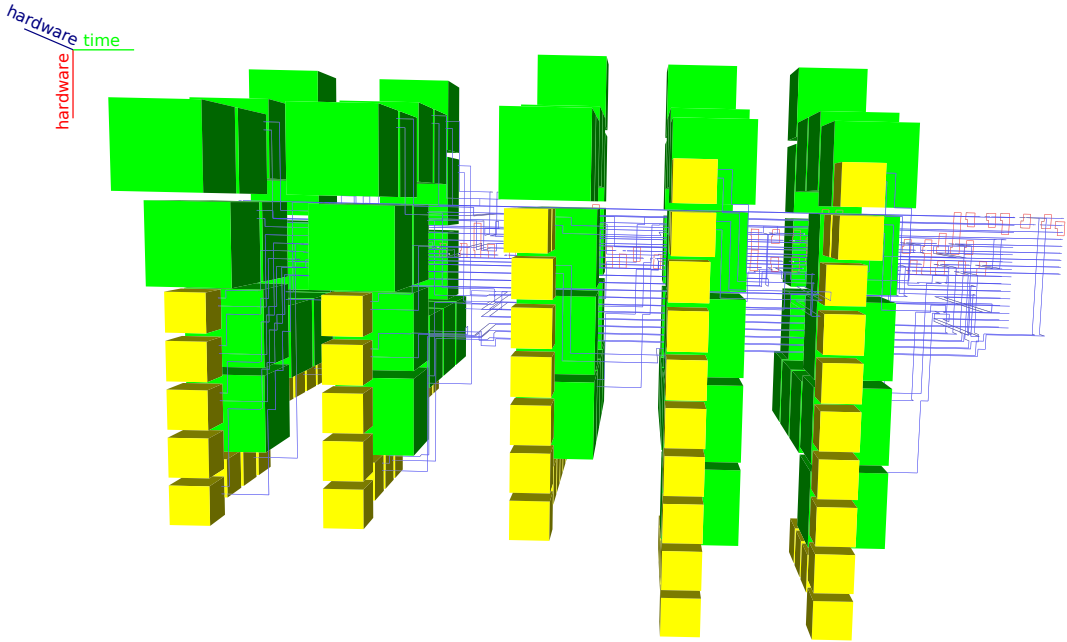}
	\caption{A synthesised assembly of topological structures protected by the surface code \cite{FMM13}. The assembly corresponds to the ICM version of a single Toffoli gate. The green and yellow boxes represent input state preparations. The 3D volume of the represented structure is an indication of the computational resources required for executing the circuit.}
	\label{fig:candy}
\end{figure}

Large scale quantum circuits are becoming state-of-the-art benchmarks for the evaluation of design automation methods scalability. Such circuits need to be quantum error-corrected, and the surface code \cite{FMM13} requires rewriting arbitrary circuits into a specific fault-tolerant form. In practice, such large fault-tolerant quantum circuits will be dynamically synthesised into assemblies similar to the one from Fig.~\ref{fig:candy}.

The data structures used by the design automation tools become increasingly important. Some tools express quantum circuits as gate lists~\cite{soeken2012revkit}, interaction graphs~\cite{anders2006fast} or assembly languages similar to QASM \cite{balensiefer2005evaluation}. Furthermore, many if not most of the underlying design automation methods rely on rewriting circuit structures (manipulating the circuit gates), as well as reordering (manipulating the position of qubits/wires). 
More precisely:
\begin{itemize}
\item \emph{Rewriting} is a quantum circuit manipulation in which certain gate sequences are replaced by other functionally equivalent gate sequences (e.g.~in order to reduce the costs).
Rewriting is e.g.~applied in~\cite{miller2003transformation,maslov2008quantum}.

\item \emph{Reordering} is a quantum circuit manipulation where the order of a circuit's wires is changed (e.g.~in order to make a quantum circuit nearest neighbour compliant). It does not influence the functionality of the circuit.
Reordering is e.g.~applied in~\cite{DBLP:journals/tcad/WilleLD14,zulehner2017exact}.

\end{itemize}

Since both operations come with a polynomial complexity, naive implementations of the corresponding rewriting and reordering schemes have been sufficient thus far. Moreover, circuit manipulations are required to maintain the initial circuit's operations partial ordering: some operations could be allowed to execute in parallel. However, the sheer number  of possibilities to consider  make these implementations highly inefficient -- constituting a serious bottleneck.

In this work we investigate the actual computational complexity of existing state-of-the-art quantum circuit manipulation with a particular focus on rewriting and reordering. It should be noted, that manipulation refers to the classically executed design procedures, and not the simulation or execution of the circuits.  We observe that -- despite the polynomial complexity in theory -- existing solutions require rather non-negligible classical computational efforts. With increasing scales of quantum computations, as well as corresponding quantum circuits, this becomes a bottleneck for various design steps. As a consequence, alternative methods for rewriting and reordering are required. 

The basic idea presented in this work is to employ diagram-like structures to reduce the computational complexity of updating/propagating structural changes (manipulations) in the circuit. Empiric evaluation results show that speed-ups of orders of magnitude (i.e.~more than 1000x) are possible by using the proposed methods. This means that the same circuit design steps can be executed a thousand times faster compared to the naive state-of-the-art implementations.

\newpage

\subsection{Quantum circuits}
\label{sec:qc}

In order to keep this paper self-contained, the quantum circuit formalism is briefly reviewed. This section covers the elements of the fault-tolerant circuits used in this work.

Quantum computations have a reversible formulation, meaning that the inputs could be computed from the outputs by applying the circuit backwards. This is possible because (1)~quantum circuit gates have an equal number of inputs and outputs and (2)~gates implement a specific type of operation. Computational reversibility is lost when quantum measurements are included in the quantum circuit. Measurements are used to control the computation (in the measurement based computational model), to control the error-correction mechanisms (e.g. for computing error syndromes) or to read computational results out.

\begin{figure}[t]
	\centering
	\includegraphics[scale=1.2]{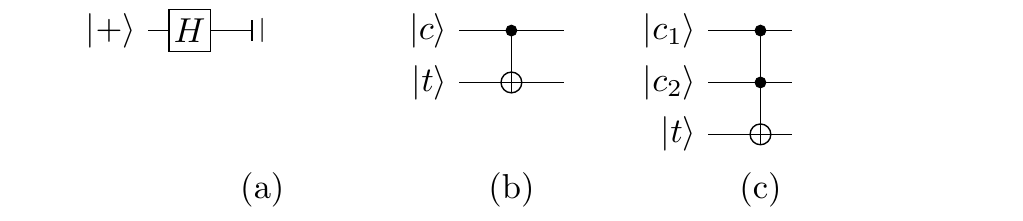}
	\caption{Circuit examples. The $\bullet$ symbol indicates control qubits, and $\oplus$ target qubits: a) a single qubit circuit having the input initialised to $\ket{+}$, operated by a gate $H$ and the qubit is finally measured; b) the CNOT gate; c) the Toffoli gate. Qubit initialisation and measurement were left out in circuits b-c).}
	\label{fig:circex}
\end{figure}

Classical Boolean circuits can be made reversible by using gates implementing bijective Boolean functions, and having an equal number of inputs and outputs as well~\cite{NC00,DBLP:conf/date/ZulehnerW17}. Classical reversible circuits do not have the computational power of their quantum counterparts, but are described using the same circuit formalism and often serve as initial blueprint~\cite{saeedi2013synthesis,shende2003synthesis,DBLP:conf/aspdac/NiemannWD14,miller2003transformation}.

A quantum circuit is a sequence of operations applied to a set of qubits. The circuit is executed by applying a gate list: an ordered set of operations, which are generally written from left to right. A time axis running left-to-right can be linked to any quantum circuit, so that each circuit operation can be identified at a certain relative point in time. Inputs are on the left and outputs on the right. Fig.~\ref{fig:circex} illustrates corresponding examples.

A quantum circuit includes three types of operations: (1)~qubit initialisation; (2)~quantum gates which operate on qubits; (3)~qubit measurement. 

In general, wire and qubit are used interchangeably, but it is possible to reuse the same wire for multiple qubits (e.g.~Section~\ref{sec:reord}). This distinction will become important in the following sections.

A circuit's execution starts with qubit initialisation. Particularly in the context of practical large scale fault-tolerant quantum computation, qubits are initialised with states $\ket{0}$, $\ket{1}$, $\ket{+}=\frac{1}{\sqrt{2}}(\ket{0}+\ket{1})$, $\ket{Y}=\frac{1}{\sqrt{2}}(\ket{0}+i\ket{1})$ and $\ket{A}=\frac{1}{\sqrt{2}}\ket{0}+e^{i\pi/4}\ket{1})$. Qubit states are manipulated by quantum gates, which
operate on single or multiple qubits. 
Quantum gates are chosen from a universal gate set sufficient for approximating any quantum computation with arbitrary precision. Currently, the Clifford+$T$ gate set receives a lot of attention, because there are known methods to implement it fault-tolerantly. This set includes the CNOT gate (Fig.~\ref{fig:circex}b). Boolean reversible circuits are generally expressed using only Toffoli \cite{NC00} gates (Fig.~\ref{fig:circex}c). The Toffoli gate is a 3-qubit gate applying the NOT operation on the target $t$ if both control qubits $c_1, c_2$ are $\ket{1}$.

Quantum measurement is the last operation applied to a qubit. The standard measurement, referred to as a $Z$-basis measurement, measures if the qubit is in the $\ket{0}$ or $\ket{1}$ state, collapsing any superposition inconsistent with the measurement result. Another type of measurement is an $X$-basis measurement, which measures if the qubit is in the $\ket{+}=\frac{1}{\sqrt{2}}\left(\ket{0}+\ket{1}\right)$ state or the $\ket{-}=\frac{1}{\sqrt{2}}\left(\ket{0}-\ket{1}\right)$ state.

\section{Methods}
\label{sec:prelim}

Rewriting and reordering represent established methods frequently applied in the design of quantum circuits. The first part of this section offers a high level perspective and illustrates the methods on circuit examples. The second part of the section introduces the proposed solutions for speeding up these circuit manipulations.

It is envisioned that the first reliable quantum computers will be based on surface quantum error-correcting codes (QECCs) \cite{van2016path}, and for practical reasons, this work focuses on the design automation of large scale fault-tolerant quantum circuits compatible with such QECCs. The following examples are mostly based on a circuit form called ICM (Initialisations, CNOTs and Measurements) \cite{paler2017fault}. Arbitrary quantum circuits should, before being executed and if surface QECCs are to be used, be compiled into this form. Such circuits are universal and include CNOT gates and all the initialisations and measurements presented in Section~\ref{sec:qc}. The methods in this section are applicable to other types of circuits too, but the focus is on ICM because of its structural simplicity and ease of illustration.

\subsection{High level perspective}

Quantum circuit manipulations are not used only for compiling equivalent forms of a circuit, but also for various circuit optimisations. Quantum circuit optimisation's task is to minimise a cost function which can take different forms ranging from number of gates \cite{saeedi2013synthesis,DBLP:conf/ismvl/MillerWD10}, number of wires \cite{DBLP:journals/integration/WilleSMD14,paler2016wire}, number of SWAP gates required to achieve linear nearest neighbour interactions \cite{hirata2011efficient,DBLP:journals/tcad/WilleLD14} or the space-time volume necessary to execute the fault-tolerant version of the circuit \cite{fowler2012time}.

The space-time cost of a quantum circuit continues to receive considerable attention \cite{fowler2012time, bishop2017quantum}, because it expresses the minimum computational requirements a quantum computer would need in order to execute the circuit. The number of wires is an indication of the space (hardware) requirements, and the number of gates (circuit depth) an expression of an execution's time cost. Mapping a quantum circuit to distinct quantum computing architectures will result in different cost models, but the analogy will still hold.

\subsubsection{Rewrite.}
\label{sec:rewr}

Rewriting can be applied to optimise circuits without changing the used gate set (e.g. replace a CNOT gate sequence with a functionally equivalent one). It is also used to transform (compile) a circuit from one universal gate set into another. Arbitrary quantum circuits can be expressed using the Clifford+$T$ gate set after decomposing each gate from the initial gate list into a Clifford+$T$ sequence (e.g.~decomposing an arbitrary single qubit Z axis rotation \cite{ross2014optimal}).

Rewriting is used also, for example, to decompose the Toffoli gate into Clifford+$T$ gates (e.g. \cite{selinger2013quantum}). There are multiple equivalent decompositions: some use ancillae (e.g. \cite{selinger2013quantum}), while others do not (e.g. \cite{NC00}). For the purpose of the following discussion, rewrites introducing ancillae are of interest, because ancillae are used as temporary workbenches and their usage can be optimised (see following section). 

Furthermore, rewriting is used for the compilation of ICM circuits, too, by further decomposing Clifford+$T$ gates into ICM primitives.

\begin{figure}[h!]
	\centering
	\includegraphics[width=.7\columnwidth]{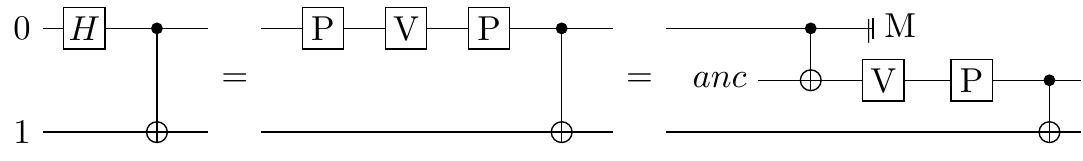}
	\caption{Circuit rewrite example: A Bell pair construction circuit is iteratively rewritten by decomposing first the H gate into a sequence of three gates. One of the resulting gates (the first P gate) is rewritten using a teleportation, such that a CNOT gate and an ancilla are included in the circuit. The circuit is not fully ICM at this point. For this the remaining V and P gates have to be decomposed, too.}
	\label{fig:ex1}
\end{figure}

\begin{example}
\label{ex:pvp}
A two qubit quantum circuit is rewritten in ICM form (see Fig.~\ref{fig:ex1}). The initial gate list is [h 0, cnot 0 1]. Certain classes of QECCs cannot implement the Hadamard gate (denoted by $h$) directly, and the gate is decomposed into a sequence, such that the resulting gate list is [p 0, v 0, p 0, cnot 0 1], where $p$ is the square root of the $Z$ gate, and $v$ is the square root of the $X$ gate. Both the $p$ and $v$ gate are implemented through teleportation in ICM. Therefore, each instance is replaced by an ancilla initialised into a specific state, a CNOT gate and a single qubit measurement. After using ancilla initialisation (init) and the measurement (meas), the first replacement will result in the gate list [init anc, cnot 0 anc, meas 0, v anc, p anc, cnot anc 1], where $anc$ is the identifier of the newly introduced ancilla.
\end{example}

\subsubsection{Reordering and reachability analysis.}
\label{sec:reord}

\begin{figure}[h!]
	\centering
	\includegraphics[scale=1]{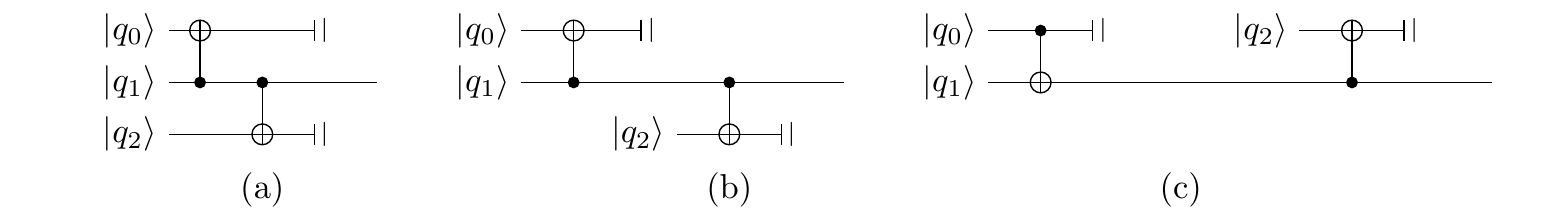}
	\caption{Wire recycling example: three wires, used for three qubits, are reduced to only two wires holding the same qubits, but the wires are ordered differently in time \cite{paler2016wire}. The difference between (a) and (b) is that the three wires have a different time ordering. In (b) the third wire could be removed, because $\ket{q_2}$ is used only after $\ket{q_0}$ is measured. Thus, the two qubit can share the same wire as shown in (c). The temporal ordering in (b) and (c) is the same.}
	\label{fig:exwr}
\end{figure}

Various quantum circuit optimisation tasks are related to reordering qubits. Moreover, qubits can be ordered in space or in time. Time optimisation focused on achieving an efficient linear nearest neighbour interaction, while space optimisation is used to reduce the total number of wires (hardware). An example of the latter is wire recycling \cite{paler2016wire}, where multiple qubits are placed on the same quantum circuit wire. This reflects a temporal ordering between the circuit's qubits (e.g. Fig.~\ref{fig:exwr}). Space and time reordering of qubits are performed after analysing the circuit's structure. 

In the case of recycling, a \emph{reachability analysis} is performed. The analysis determines which qubit measurements (outputs) are \emph{reached} from each qubit initialisation (inputs). Inputs, from which a particular output is not reached, can be placed in the circuit after that output. This means, that the initialisations can be placed on the same wire after the measurement, because those initialisations do not influence the preceding measurement.

\begin{figure}[h!]
	\centering
	\includegraphics[width=0.9\columnwidth]{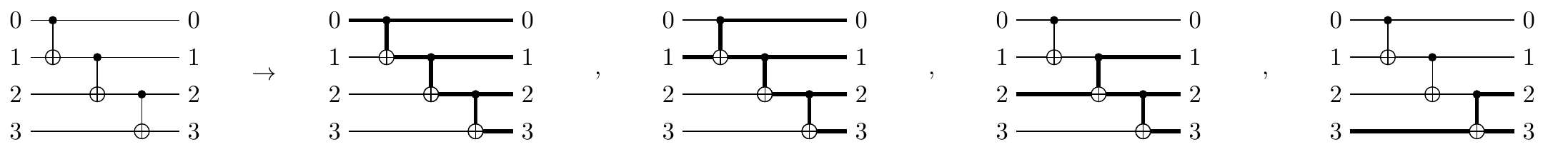}
	\caption{Reachability analysis example: A four qubit circuit consisting of three CNOT gates, and the four results of the reachability analysis starting from each input.}
	\label{fig:ex2}
\end{figure}

\begin{figure}[h!]
	\centering
	\includegraphics[width=0.7\columnwidth]{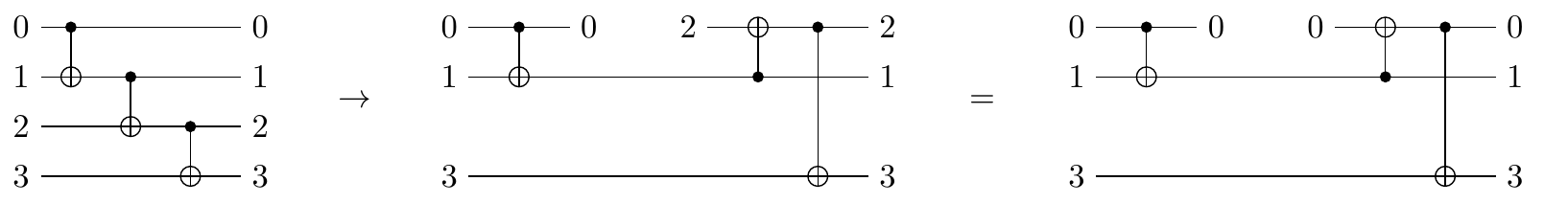}
	\caption{Wire recycling example: A four qubit circuit consisting of three CNOT gates has wires 0 and 2 recycled. The resulting circuit has three wires. After recycling and updating the references (2 is updated to 0), the circuit operates on the wires 0, 1 and 3.}
	\label{fig:ex3}
\end{figure}

\begin{example}
\label{ex:2}
Consider a four qubit quantum circuit consisting of three CNOT gates structured as [init 0\ldots 3, cnot 0 1, cnot 1 2, cnot 2 3, meas 0\ldots 3] (see Fig.~\ref{fig:ex2}). The initialisation of qubits 0 and 1 has the potential to influence the outputs of qubits 0, 1, 2 and 3. Qubit 2 initialisation reaches the outputs of qubits 1,2 and 3. The initialisation of qubit 3 reaches only the outputs of qubits 2 and 3. The structure of the circuit was considered fixed, such that gates are not commuted. Commuting gates could generate different sets of reached outputs.

The previous reachability analysis shows that, for example, qubit 2 initialisation could be performed after the measurement of qubit 0. As a result, if the initial circuit operates on four wires (0\ldots 3) each occupied by the four qubits (0\ldots 3), after a first recycling round wire 0 will be used by both qubit 0 and qubit 2.

Recycling qubit 0 and 2 (see Fig.~\ref{fig:ex3}) on the same wire will result in the gate list [init 0, cnot 0 1, meas 0, init 1 \ldots 3, cnot 1 0, cnot 0 3, meas 0,1,3]. Having recycled wire 0, all the gates have to reference the new wire (0) instead of the old one (2). For this reason, cnot 1 2 becomes cnot 1 0, and cnot 2 3 becomes cnot 0 3, while meas 2 becomes meas 0.
\end{example}

\subsubsection{Rewrite vs. Reorder - When ancillae are costly to manipulate}

There are the two scenarios regarding the structure of a quantum circuit: a) gates are ordered, but wires not; b) both gates and wires are ordered. Simultaneously, the following distinction between rewriting and reordering has to be noted: rewriting has the potential to change both the number of gates and qubits/wires of a circuit, while reordering does not influence the number of qubits and gates. Reordering introduces/changes the temporal ordering of the qubits, and it can lead to changes in the number of wires (as seen in wire recycling).

There are two rewrite situations: either ancilla qubits (wires) are introduced into the circuit, or not. The first situation may result in slow manipulations, because wire labels may need to be updated. The latter is not so complex. In practice, ancillae are introduced when expressing a gate through an equivalent sub-circuit (e.g. ICM and Toffoli decompositions in Section~\ref{sec:rewr}). For the ICM case, fault-tolerant single qubit gates are implemented through teleportation mechanisms which require at least a single ancilla, the use of single qubit measurement, single qubit initialisation, and CNOT gates. Introducing a single ancilla is equivalent to \emph{splitting a wire in two}.

Reordering wires is, in certain cases, the inverse operation of rewriting where ancillae are introduced: \emph{wires are joined}, meaning that the same wire is reused by multiple qubits. This happens during wire recycling, which is motivated by the fact that the hardware requirements of a large scale quantum circuit can be reduced by reusing the same wire for multiple qubits.  Nevertheless, qubit reordering can affect the number of SWAP gates required to execute the circuit on quantum computer architectures (chips) with limited connectivity. 

In conclusion, rewrite may split an existing wire into two distinct wires, whereas recycling joins two wires into a single one.

\subsubsection{Random access manipulations.}

There is a generally accepted workflow of how quantum circuits are manipulated. Rewrites are the first to be applied in order to transform the input circuit into a gate set compatible with the underlying quantum computing architecture. Reorderings are used for layout optimisation of the resulting rewritten circuits. Therefore, it is assumed that the rewrites and reorderings are not applied randomly (in the sense of RAM, and not stochasticity).

This work argues that design automation tools should be able to apply manipulation, irrespective of their types, randomly. Circuit compilation, optimisation and execution \emph{will not form a linear workflow} (one follows the other) but a circular one (execution results will influence the next compilation and optimisation). The circular workflow may be applied in an online or offline manner. The first appears when computations are actually executed on real quantum computers. The latter exists when quantum circuits are designed before being executed.

The online circular workflow will be used within the control software of large scale quantum computers. This workflow requires: (1)~random access to the quantum circuit being executed, (2)~eliminating unnecessary updates to the input circuit. The requirements are discussed using a scenario which imagines a situation where gates can be conditionally inserted into the gate list but can't be run directly, because they need to be compiled (as shown in the following).

A quantum computation's execution starts from a non-fault-tolerant Clifford+$T$ circuit. The first step is to compile the input circuit to a fault-tolerant circuit, which consists of QECC structures (e.g. surface code). However, even though the circuit is error-corrected and fault-tolerant, the execution of certain gates is not deterministic. For example, this is the case of the $T/T^\dagger$ gates which may need the corrective application of $S$ gates. There are two possibilities: (1)~either the $S$ gates are always compiled into the gate list before executing the circuit, but applied only when needed (simple situation), or (2)~the gates are dynamically inserted into the gate list each time they are required (complex situation). The first option, compared to the latter, is more expensive in terms of computational resources after the error corrected quantum ciruits are compiled. Resource efficiency is one of the most important aspects during fault-tolerant circuit compilation, such that the complex situation is encountered in practice.

The online circular workflow includes a feedback loop between quantum hardware and the input non-fault-tolerant Clifford+$T$ circuit. Quantum computation is halted whenever non-deterministic $T/T^\dagger$ gate performs the undesired operation (determined based on signals received from the hardware), and the necessary $S$ correction gate is introduced into the gate list of the input Clifford+$T$ circuit. The workflow's complexity increases additionally if quantum circuit optimisations are dynamically performed based on the newly inserted correction gates. Circuit execution uses a linear traversal of the initial gate list and gate corrections are inserted into the list right after the currently executed gate, but before any other gate that will be executed. However, the \emph{control software cannot execute correction gates directly}, because the corrections also need to be compiled and optimised into QECC structures. 

A data structure allowing random access and lazy updates is useful in the online workflow, because: (1)~gate insertions can influence the next gates from the list (gate type, wires operated on), and those gates have to be easily accessed instead of iterating each time to the end of the gate list, (2)~multiple updates may be unnecessary (e.g. the same gate is updated to operate on wire $w1$, afterwards on wire $w2$ etc. and until this gate is executed the gate operates on wire $wn$).

The offline circular workflow is mostly related to quantum circuit optimisation, where the goal is to use as few resources as possible to implement a given quantum computation. Realistically, brute force approaches do not scale for large circuits, such that heuristics are the methods of choice. Heuristics, as reviewed for example in \cite{saeedi2013synthesis}, include transformation rules and template-based optimisation, which operate locally on the circuit. More specifically there are heuristics which select a subcircuit and apply various optimisation methods to it. It can happen that gate optimisations are followed by methods which add additional wires (qubits) for optimisation purposes etc. A heuristic works in multiple passes (may include also a few backtracking steps when modelled as a search problem \cite{saeedi2013synthesis}). 

The offline circular workflow takes an input quantum circuit and iteratively transforms into a more resource efficient quantum circuit. Multiple linear traversals (passes) of the circuit are repeated and circuit manipulations can be organised into a hierarchical data structure: the number of the pass is an indication of the hierarchy level (e.g. a tree where the first pass is at the highest level, and last pass at the lowest level/leafs). Such a data structure is used to \emph{collect} multiple distinct manipulations into a single final manipulation, which is applied in a lazy manner to the input circuit. The advantage is that unnecessary circuit updates are eliminated. The disadvantage is that a random access model would be necessary, because: (1)~a single update operation abstracts a sequence of hierarchically organised updates which were not applied, (2)~ the single update has to be easily located in the input circuit, (3)~it cannot be assumed that updates are performed respecting the order of the input circuit, (4)~it is faster to traverse the hierarchical updates data structure linearly and to apply the collected updates randomly to the circuit, than to traverse the input circuit linearly and to determine on a gate-by-gate basis which collected updates are necessary.

Consequently, in order to speed circuit manipulations, a novel circuit manipulation data structure has a hierarchical organisation and supports interleaved random access manipulations of quantum circuits.

\subsubsection{Complexity - The naive scenario}

Reordering and rewriting do not have a high computational complexity. For a circuit of $q$ qubits and $g$ gates, \emph{the naive implementation of any manipulation} will have a worst-case computational complexity of $\mathcal{O}(qg)$: after each manipulation the entire circuit is updated. 

Not all circuit manipulations update wire labels from the gates, such as the Hadamard decomposition from the first example, and the Toffoli gate decomposition from \cite{barenco1995elementary}. However, in the worst case the manipulation of all the gates from a list is $\mathcal{O}(qg^2)$: each gate could trigger updates in all other $g$ gates (therefore $g^2$), which in the worst case operate on the maximum number of wires $q$ (resulting $qg^2$).

Although the update complexity is polynomial, the execution times are very slow when manipulations are implemented naively (as evaluated by experimental evaluations summarised later in Section~\ref{sec:res}). In the following, we call \emph{naive methods} the simplistic implementation of the worst case manipulation scenario: update all the gates after each manipulation which affects the wires. The issue is that the naive methods do not recognise if the same update is performed repeatedly for multiple gates, or if some updates are not necessary because they will be made obsolete by a future manipulation.

This motivates more efficient and elaborated methods for executing quantum circuit manipulations. The focus will be on rewriting and reordering, which are introduced in the following sections, because these are necessary for preparing large quantum circuits. The herein presented methods will offer significant speed-ups, because tree like structures (wire label reference diagrams) are used to represent structural decisions taken at each circuit gate (cf. if-then-else). Each circuit manipulation inserts a decision into the diagram: how is the new wire related to the pre-manipulation structure?

\subsection{Wire label reference diagrams}
\label{sec:def}

Quantum circuits, as used in the design automation community, consider wire labels an indication of ordering, and not a reference. This differs from the mathematical formalism underlying quantum circuits, where $\ket{q_0q_1q_2}$ and $\ket{q_1q_0q_2}$ are conceptually the same state but with reordered qubits.

Faster circuit manipulation can be achieved by treating qubit labels as references and not order indicators.

From a symbolic perspective, the gates $I_0\otimes CNOT_{1,2}$ can be applied to both $\ket{q_0q_1q_2}$ and $\ket{q_1q_0q_2}$, because both the gates (e.g. CNOT) and the qubits use \emph{references} and not numbers expressing a strict order (e.g. position in a list). In other words, for example, the gate $h$ $1$ is not the application of the Hadamard gate on the first qubit, but on the qubit having the label $1$.

A circuit where no manipulations were performed consists of a gate list, which is an ordered sequence of operations applied to qubits. 

\begin{definition}
A \emph{gate identifier} indicates the position of the gate in the ordered gate list.
\end{definition}

\begin{figure}[h!]
	\centering
	\includegraphics[scale=1.0]{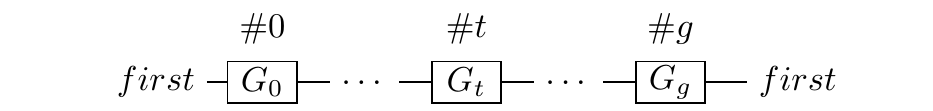}
	\caption{A single wire circuit consisting of $g+1$ arbitrary single qubit gates. Gate identifiers are represented above the gates.}
	\label{fig:construct0}
\end{figure}

Integer numbers are the simplest implementation of gate identifiers (indicated in this work by the prefix \#). For example, in the ordered gate list $[p, v, t]$ the gates are identified by their order number: the gate $p$ by \#1, the gate $v$ by \#2, and the gate $t$ by \#3.  Fig.~\ref{fig:construct0} illustrates the identifiers in a single-qubit circuit. A general type of identifiers is presented in Section~\ref{sec:ids}.

\begin{definition}
A \emph{wire identifier} is the label (name) associated to the qubit, where a label does not indicate any existing ordering (e.g. between qubits).
\end{definition}

Each wire is a distinct object, thus not related in any way to the other wires. Returning to the state $\ket{q_0q_1q_2}$, the three qubits are identified by their names $q_0$, $q_1$ and $q_3$. The subscripts do not indicate an ordering, e.g. the $2$ in $q_2$ is just part of the name, and does mean that this particular qubit is the second one in any list. At this point, it is possible to define wire label references.

\begin{definition}
A \emph{wire label reference} is an indicator to a wire identifier.
\end{definition}

Consequently, from a data modelling perspective, each circuit operation has: (1)~a specific identifier, and (2)~a list of references to the wires it operates on. Initialisations, measurements and gates are circuit operations and are, consequently, modelled similarly.

\begin{figure}[h!]
	\centering
	\includegraphics[width=0.4\columnwidth]{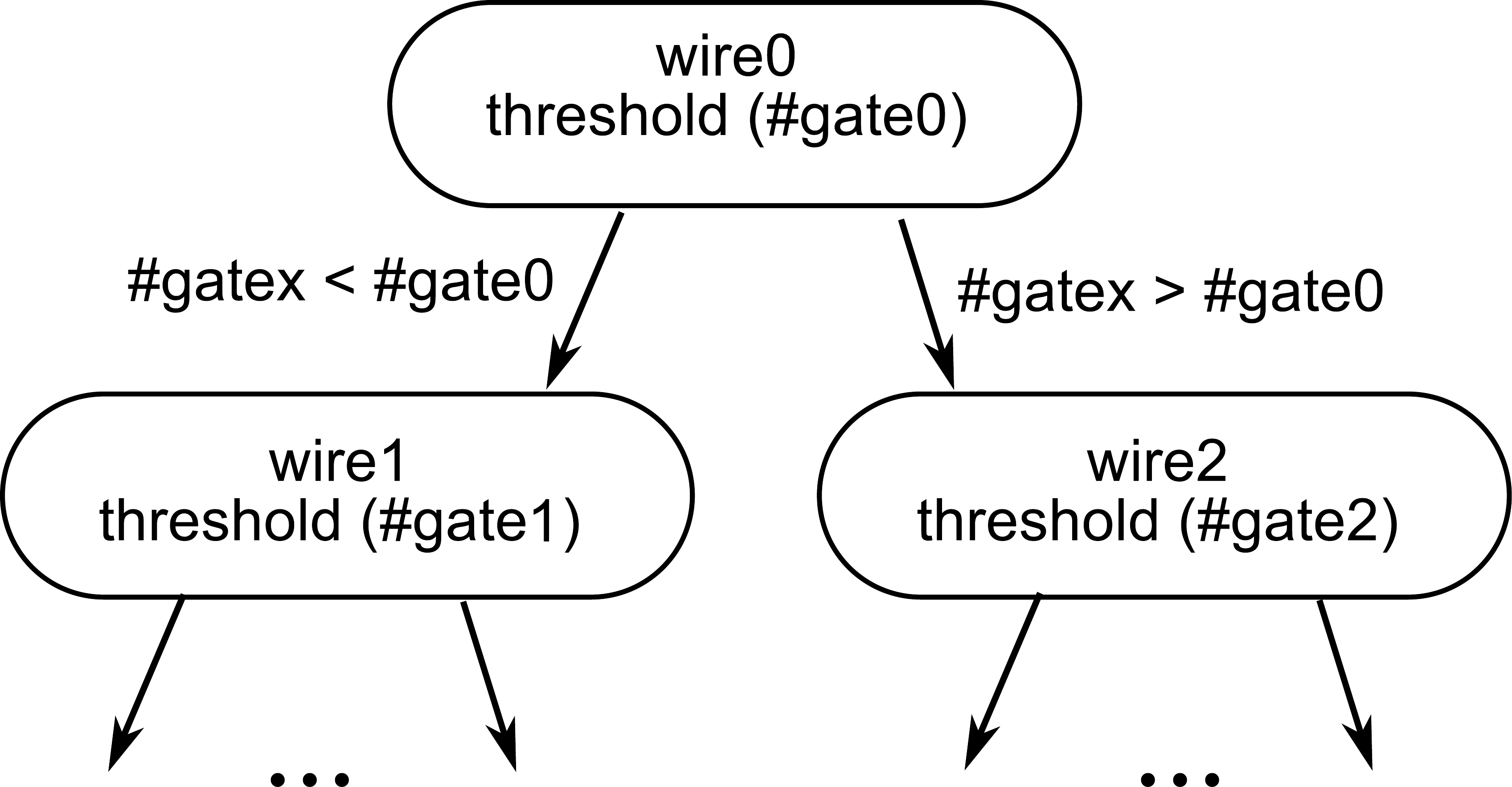}
	\caption{A wire label reference diagram.}
	\label{fig:diag1}
\end{figure}

\begin{definition}
A \emph{wire label reference diagram} (e.g. Fig.~\ref{fig:diag1}) is a tree with nodes representing label references. Nodes and edges are added to the diagram each time a manipulation affects the wires (split or joined).
\end{definition}

\begin{definition}
A \emph{diagram node} (e.g. there are three in Fig.~\ref{fig:diag1}) has at most two children, and contains a label reference (e.g. wire0) and a \emph{threshold} value (e.g. \#gate0).
\end{definition}

\begin{definition}
The \emph{threshold} value is a gate identifier used to determine at which position in the circuit there is a decision to be taken about which wire to use.
\end{definition}

The per node threshold value is related to the gate identifiers and their interpretation as indicators of the operation ordering. For a circuit represented as a sequence of operations $G_0, \ldots, G_t, \ldots G_g$, the threshold $t$ indicates that after manipulating operation $t$, the ones preceding $t$ will reference a different wire than the gates succeeding $t$.

\begin{definition}
\emph{Diagram traversals} are functions of gate identifiers, and are used for updating the label references associated with each gate.
\end{definition}

The application of such diagrams is exemplified using manipulations representative for large scale fault-tolerant quantum circuits: ICM rewriting and ICM recycling. The construction and usage of the diagrams will be introduced next.

\subsubsection{Rewrite: ancillae are introduced.}

The effect of rewrites on a diagram can be visualised starting from a circuit having a single wire and a $(g+1)$-long sequence of arbitrary single qubit gates $G$. The initial diagram will include a single node. 

\begin{figure}[h!]
	\centering
	\includegraphics[width=0.6\columnwidth]{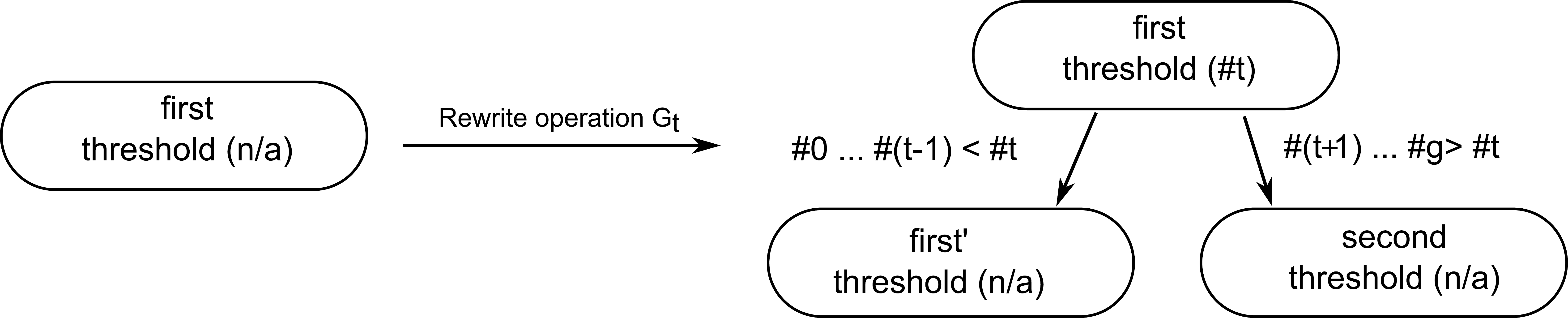}
	\caption{The initial circuit consists of a single wire. After rewriting an operation the diagram is updated with two nodes.}
	\label{fig:simplu}
\end{figure}

Rewriting gate $G_t$ splits the initial wire (called first) into two distinct wires (first$'$ and second). The wire label diagram resulting after the rewrite is presented in Fig.~\ref{fig:simplu}. The root node of the figure has the threshold set to $\#t$. The threshold property of the diagram leaf nodes (first$'$ and second) is not set (n/a).

The new sub-wires are associated to two new diagram nodes appended as children of the initial node. The children are called \emph{left node}, e.g. first$'$ in Fig.~\ref{fig:simplu}, and \emph{right node}, e.g. second in Fig.~\ref{fig:simplu}. The circuit operations before $t$ will need to reference the left node, and the  operations after $t$ the right node.

The rewrite operation inserts for the new gates the correct wire label references automatically. However, all other gates will require a reference update. From the initial, the preceding (with identifiers smaller than \#t-1) and succeeding (identifiers higher than \#t+1) gates still reference the wire first, whose threshold value is set to \#t.

If one of the sub-wires (e.g. child nodes first$'$ or second) is split in a next step, then their threshold parameter is configured to the gate number that generated the split. The manner in which new operation identifiers are introduced influences future threshold values. These details are discussed in Sections~\ref{sec:ids}.

\subsubsection{Reordering: wires are reduced (recycled).}

\begin{figure}
	\centering
	\includegraphics[width=0.7\columnwidth]{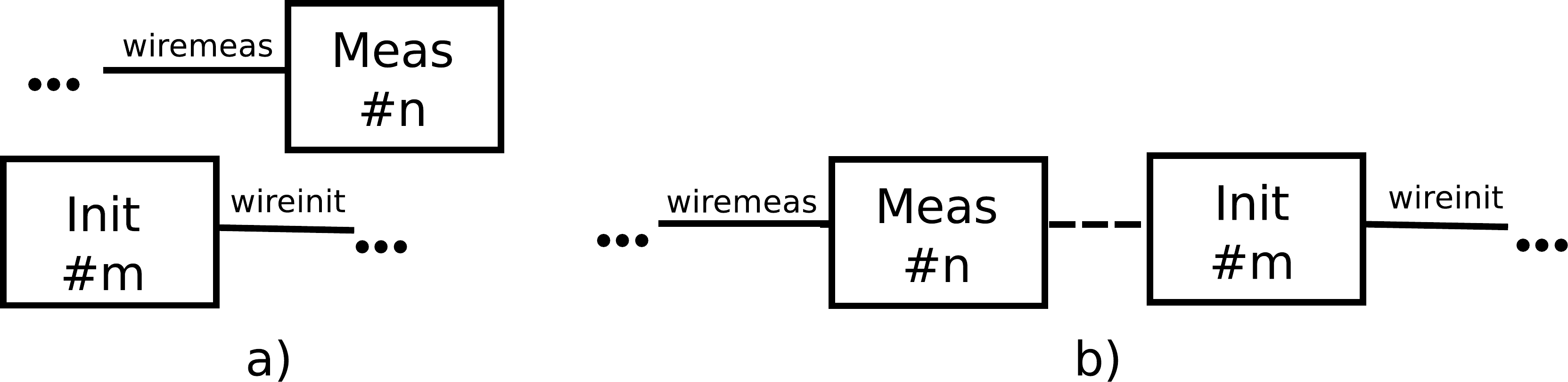}
	\caption{Recycling example: a) the initialisation and measurements operate on different wires; b) initialisation and measurement are on the same wire.}
	\label{fig:diag4}
\end{figure}

A recycling example is illustrated in Fig.~\ref{fig:diag4}, where an initialisation (operating on the qubit labelled as wireinit) can be placed on the same wire after a measurement (labelled as wiremeas). The reachability analysis (more details in Section~\ref{sec:reach}) indicated that wiremeas is not reached from wireinit, such that the two wires are recycled.

It could happen that in the original gate list, the initialisation operation is numbered \#m, and the measurement \#n, such that the circuit diagram will represent the initialisation earlier than the measurement ($\#m < \#n$). However, after recycling the new order will be such that $\#m > \#n$, because the initialisation is placed after the measurement. This is possible because of the result of the reachability analysis: the initialisation does not influence the state measured. This shows the partial ordering existing between the circuit operations: in fact the initialisation and the measurement could have been executed in parallel, but in order to save hardware resources are executed sequentially and share the same wire.

\begin{figure}
	\centering
	\includegraphics[width=0.4\columnwidth]{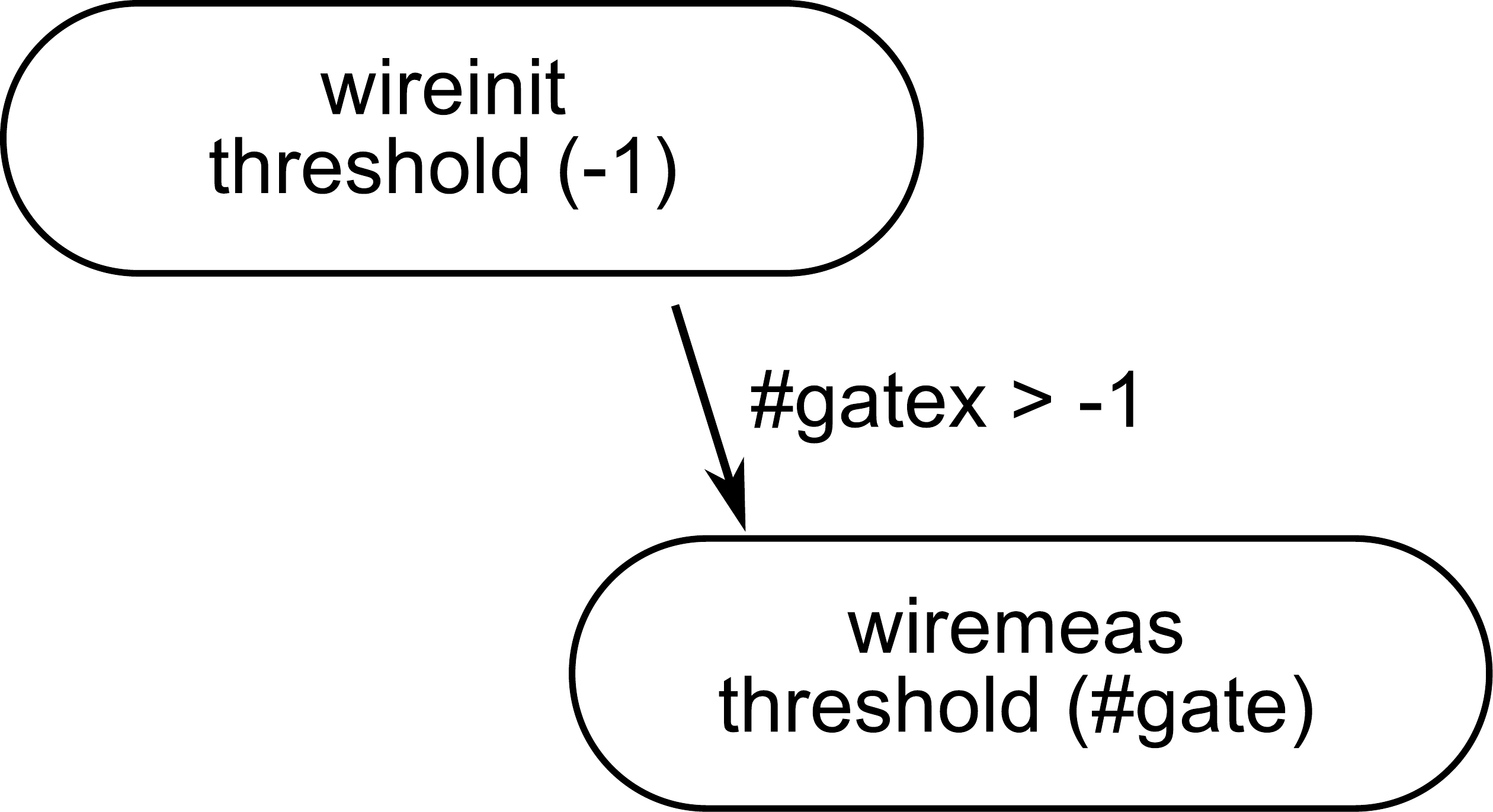}
	\caption{Wire label reference diagram resulting after the recycling from Fig.~\ref{fig:diag4}.}
	\label{fig:diag2}
\end{figure}

The wire label reference diagram resulting after the recycling example from Fig.~\ref{fig:diag4} is shown in Fig.~\ref{fig:diag2}. All the gates that used wireinit (the wire of the initialisation) will reference the wire of the measurement (wiremeas). The wiremeas diagram node is connected as a right child of the wireinit node: the reference of the measurement wire is reused (recycled) for initialisation. The gates following the initialisation will have to update their label reference (when needed), and the update process is a search controlled by the threshold node parameter. Because all gates have an identifier higher than -1, the threshold of the initialisation wire can be safely set to $-1$: all gates will automatically receive the updated reference.

\subsubsection{Constructing and using the diagrams.}
\label{sec:speed}

Wire label reference diagrams circumvent the naive update of wire labels. In a circuit with $g$ gates, each manipulation is generally applied on a single gate, but its effect on the wire labels has to be propagated to the other $g-1$ gates. Manipulating an entire circuit requires in the worst case $g(g-1)$ operation updates.

By using wire label diagrams, each operation can infer its wire labels only when needed. This saves a large number of \emph{unnecessary} updates, because \emph{eager} label updates are replaced by \emph{lazy updates}: the most up to date wire labels are determined before manipulations are performed. The circuit's structural changes are stored in the diagram, a binary tree, which is iteratively updated after each circuit's structural change. If multiple manipulations have affected the wire labels of a certain gate, the gate determines its most up to date wire label through a tree search. 

The following is an illustration of Example~\ref{ex:pvp}, the construction of a Bell pair. Two rounds of rewrites are performed, followed by a recycling round. Operation identifiers are integers (a simplification of the solution proposed in Section~\ref{sec:ids}), and updates are forced after each round.

\begin{example}
The initial circuit consists of two gates (numbered 1 and 2) and two wires labelled wire0 and wireB in the diagram. The first rewrite replaces the H gate with the P,V,P sequence. Being the results of a rewrite, the three gates inherit the identifier 1. The reference diagram is unchanged. The first round of rewrites is finished, all (if any) gates are updating their label references and the gate list is renumbered.

The second rewrite round replaces the V and P gates with equivalent teleportation-based sub-circuits. These are in the ICM form, and use the operations described in Section~\ref{sec:qc}. The three rewrites will generate nine obligatory wire updates: there are three elements in each ICM sub-circuit, and their wire is determined after placing them in the circuit.

The first rewrite splits wire0 into wire1 and wire2. For example, the Z basis measurement is on wire1, and the $\ket{Y}$ initialisation on wire2. If no diagram would be used (and labels are order identifiers), because wire2 is introduced between wire1 and wireB, the operations V and P are moved from the first to the second wire. Therefore, beside updating the newly introduced ICM sequence, in a naive scenario four wires would be updated (two in the remaining V and P, and two in the last CNOT). Similarly, the second rewrite splits wire2 into wire3 and wire4. If eager updates are performed, the last CNOT would generate two additional updates. The third and last rewrite decomposes the V gate: wire3 is split into wire5 and wire6. The CNOTs numbered 3 and 4, an initialisation and a measurement (both numbered 3) need to be updated. Once more, six additional updates are necessary.
\end{example}

Besides the wire updates incurred by the decompositions would require 4+2+6=12 additional updates. The fourth CNOT (the last one) would be updated three times, because the position of wireB is constantly changed after each rewrite. Such updates are unnecessary if the diagram is used: it can be seen that wireB is not affected by any of the rewrites. The speed-up presented in the Results section stems from eliminating unnecessary updates.

\begin{example}
(Continuation of the previous one) The potential unnecessary updates can be illustrated also by the recycling round. The first recycle joins wire1 and wire6. Thus, all operations affecting wire6, wire5, wire4 and wireB would need to be updated. Unnecessary updates would be generated also after the second recycle, when wire5 and wire4 are joined.
\end{example}

Although this example did not illustrate this, diagrams can be used in a random access mode when rewrites and reorderings are performed in an arbitrary sequence. This usage would require more general gate identifiers (cf. Section~\ref{sec:ids}).

\subsubsection{Reference updates.}

Each quantum gate will have to update its list of label references. The update will take place either before the operation is manipulated or, at the latest, when the entire circuit manipulation is finished. The update process is based on a diagram traversal controlled by the threshold values stored per node. This process is analogue to a tree search: a key value is required for finding a specific value in the tree. Wire labels are the sought values. The keys for searching label references in the diagram are the operation identifiers.

Wire labels are updated by traversing the diagram from the per gate referenced nodes towards the correct diagram leaf node. Assume that a gate, which references wire0, identified as \#gatex, wants to determine the actual wire it is operating on. The gate queries the label reference (assume the top node from Fig.~\ref{fig:diag1}, by sending its number (identifier). A diagram traversal is started and, during this process, the nodes are chosen based on how the per node thresholds compare to the \#gatex value. If the threshold is higher than \#gatex (e.g. $\#gate0 > \#gatex$), the left child is chosen. Otherwise, the right child is chosen ($\#gate0 < \#gatex$). The search ends when the reached node has no threshold value set (n/a), or no children. For Fig.~\ref{fig:diag1}, the search stops if the thresholds \#gate1 and \#gate2 would equal n/a.

Diagram searches do not have a constant time cost, but one that depends on the length of the path between the node referenced by an operation and the most up-to-date reference node. However, this cost is smaller than the amount of unnecessary updates.

\subsubsection{Identifiers of rewritten operations.}
\label{sec:ids}

Each gate identifier is related to the position of the gate in the circuit's ordered gate list. It was mentioned that it is possible to consider these identifiers as numbers (integers). However, there are disadvantages by doing this. This section introduces a general format of the identifiers, but in order to motivate it the issue of using numbers is first discussed.

Rewrite operations, which introduce ancilla, introduce operations into the circuit, too. The gate list is already ordered and the operations are numbered. It would seem that the entire circuit's operation sequence and diagram threshold node values  need to be updated. This is not always necessary, because the new operations can take the number of the rewritten gate. This is due to the way the diagrams are used, and is shown in the following.

Assume once more that a single wire circuit is described by the $g+1$ operations $G_0, \ldots, G_t, \ldots G_g$ (Fig.~\ref{fig:construct0}), and operation $t$ is rewritten by a $v+1$-long sequence $R_{x0}, \ldots, R_{xv}$. After the rewrite the circuit will be $G_0, \ldots, R_{x0}, \ldots, R_{xv}, \ldots G_g$. All $G$ operations will still hold references to the initial wire node, whose threshold is set to $t$. Let this wire node be denoted by \emph{first}. The circuit will consist of two wires, and the $R$ operations will reference, by construction, one or both children of \emph{first}.

\begin{figure}[h!]
	\centering
	\includegraphics[scale=1.0]{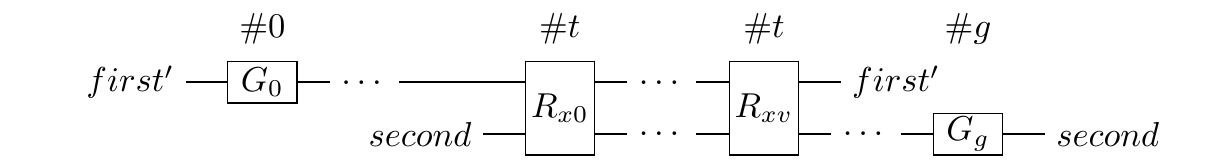}
	\caption{Gate $G_t$ is rewritten using a sequence of gates called $R$. The wire \emph{first} is split into \emph{first$'$} and \emph{second}.}
	\label{fig:construct1}
\end{figure}

It is possible to use the $t$ identifier for all the $R$ operations, because all the $G$ operations have an identifier different from $t$ (either larger or smaller, but not equal). As a result, if one of the $G$-operations would update its references, it would point also to one of the children of \emph{first}. This shows that, after a first round of initial circuit rewrites, the operation sequence does not need to be renumbered.

\begin{figure}[h!]
	\centering
	\includegraphics[scale=1.0]{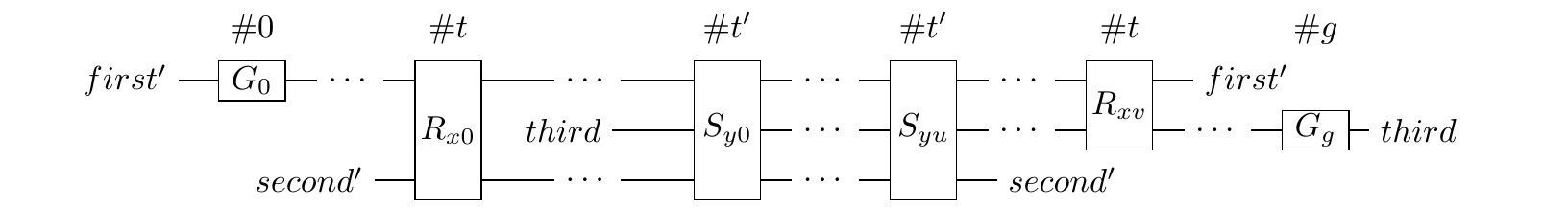}
	\caption{Gate $R_x$ is rewritten using a sequence of gates called $S$. The wire \emph{second} is split into \emph{second'} and \emph{third}.}
	\label{fig:construct2}
\end{figure}

The identifier cannot be kept (e.g. $t$, in Fig.~\ref{fig:construct2} marked $t'$) if a second round of rewrites is performed. Consider that a second rewrite affects one of the $R$ operations and not the $G$ operations (otherwise the scenario would be similar to the previous rewrite). The operation $R_{x}$ is rewritten, and introduces the $u+1$-long sequence $S_{y0}, \ldots, S_{yu}$, such that the resulting circuit is $G_0, \ldots, R_{x0}, \ldots, R_{x-1}, S_{y0}, \ldots, S_{yu}, R_{x+1}, \ldots, R_{xu}, \ldots, G_g$. The wire split by this rewrite is \emph{second}, one of the children of \emph{first}. The circuit will have three wires, all $G$ operations reference the node \emph{first}, all $R$ operations the children of \emph{first}, and all the $S$ operations children of \emph{second} (e.g. \emph{third}).

If the newly introduced operations can inherit the identifier of the original operation, it implies that $t = x$, because $x0 = \ldots = xv = t$. Furthermore, $y0=\ldots = yu = x = t$. The $S$ operations do not need to update their label references, because by construction they inherit correct children of the wire being split.

The updates of the $R$ operations will use the value $t$ to compare with the node thresholds. However, it is not clear which of the $R$ operations are before or after $R_{x}$ (thus, before or after the sequence of $S$ gates), because all have the same identifier ($t$). For the example from Fig.~\ref{fig:construct2}, it will not be clear which $R$ gates will reference \emph{second'} or \emph{third}. As a result, before the second round of rewrites the gate list needs to be renumbered (operation identifiers recomputed - introduces unnecessary updates, which should be avoided), or another identifier format is required.

\begin{figure}[h!]
	\centering
	\includegraphics[scale=1.0]{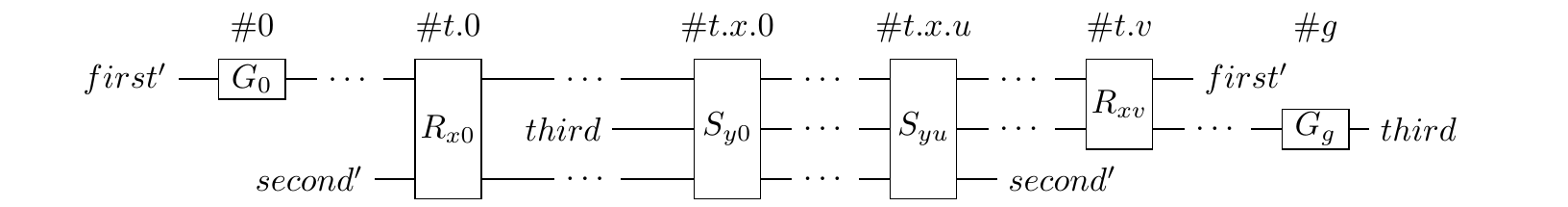}
	\caption{Using sequence-based identifiers of the form $nr_0 . nr_1 . \ldots . nr_k$ instead of integers.}
	\label{fig:construct3}
\end{figure}

The solution to circumventing the renumbering of the operation sequence is to use gate identifiers of the form $nr_0 . nr_1 . \ldots . nr_k$, where $nr$ is an operation sequence identifier and $k$ the number of rewrite rounds. The format allows unambiguous comparisons and resembles the sequence-based identifiers used in software versioning (e.g. version $3.1.7$, where $nr_0=3$, $nr_1=1$ and $nr_2=7$ for $k=2$). For example, if the initial circuit has 10 operations ($g + 1 = 10$), $t=2$ (the third operation is rewritten first), and the $R$ sequence has 11 operations ($xv + 1 = 11$), then the identifiers of the $R$ operations would be $2.i$ for $0 \leq i < 11$. After the second rewrite, if, for example $x=5$ and the $S$ sequence has 7 operations ($yu + 1 = 7$), the identifiers of the $S$ operations would be $2.5.j$ for $0 \leq j < 7$. If the general identifier format is not used, gate references have to be forcefully updated after each rewrite round.

\subsubsection{Reachability analysis using bit arrays.}
\label{sec:reach}

Wire recycling is executed after a reachability analysis was performed (e.g. Example~\ref{ex:2}). For large scale circuits, determining the sets of qubit measurements reached from each qubit initialisation can easily require a large amount of memory. For this reason, at this point, wire labels should not be considered references but numbers indicating a certain wire ordering.

This transition enables representing the sets of reached outputs as bits of an array: 1 indicates that a certain wire measurement is reached. Storing the entire reachability information requires $q^2$ bits for a quantum circuit of $q$ wires. Therefore, for approximately $10^5$ wires, about 1.25 Gbytes of memory are necessary.

\begin{figure}
	\centering
	\includegraphics[width=0.3\columnwidth]{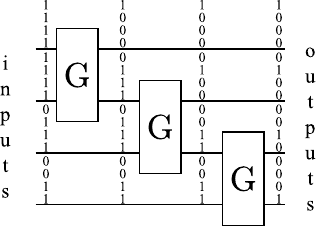}
	\caption{Reachability analysis example: the circuit is processed backwards from outputs towards inputs. A bit array is initialised with ones on the position of the corresponding wire. The arrays are updated, each time a multiple-wire gate is encountered, by OR-ing the bit arrays of the incident wires.}
	\label{fig:diag5}
\end{figure}

The reachability analysis from Fig.~\ref{fig:diag5} is the representation of Example~\ref{ex:2}. For each wire a bit array is initialised at all the positions with zero, and the position indicating the current wire with one. The bit arrays are updated backwards (from outputs towards inputs), and all the wires operated by the same gate are updated to the bit-wise OR of their respective bit arrays. For example, after the right-most gate in Fig.~\ref{fig:diag5}, the bit arrays are $0011=0010 | 0001$.

\section{Results}
\label{sec:res}

The proposed methods based on reference diagrams were implemented in the software framework used for the synthesis of topologically quantum error-corrected circuits (available at github.com/alexandrupaler/tqec/).

\begin{table*}[t]
\centering
\tiny
\begin{tabular}{l||rr|rrr||rr|rrr}
Circuit & Wires & Gates & NaiveT & FastT & ImprT & ICM Wires & ICM Ops & NaiveR & FastR & ImprR\\
\hline
add4&10&44&0.039&0.004&9.853&304&1010&0.542&0.038&14.207\\
add5&12&55&0.070&0.004&18.501&390&1297&0.980&0.055&17.808\\
add6&14&66&0.098&0.006&16.012&476&1584&1.613&0.077&21.015\\
add7&16&77&0.133&0.008&17.129&562&1871&2.318&0.102&22.740\\
add8&18&88&0.171&0.007&24.666&648&2158&3.217&0.129&25.018\\
add9&20&99&0.207&0.008&25.444&734&2445&4.201&0.159&26.442\\
add10&22&110&0.257&0.011&24.137&820&2732&5.696&0.192&29.712\\
add11&24&121&0.312&0.013&24.859&906&3019&6.864&0.227&30.204\\
add12&26&132&0.376&0.013&29.279&992&3306&8.535&0.266&32.056\\
add13&28&143&0.446&0.015&29.478&1078&3593&10.870&0.307&35.394\\
add14&30&154&0.511&0.015&33.913&1164&3880&12.289&0.352&34.907\\
add15&32&165&0.602&0.017&34.840&1250&4167&14.526&0.399&36.385\\
add16&34&176&0.758&0.019&40.708&1336&4454&17.122&0.448&38.198\\
add17&36&187&0.796&0.020&40.007&1422&4741&19.752&0.501&39.460\\
add18&38&198&0.939&0.020&47.308&1508&5028&21.953&0.556&39.513\\
add19&40&209&1.001&0.016&62.437&1594&5315&25.346&0.617&41.112\\
add20&42&220&1.143&0.024&48.514&1680&5602&28.909&0.678&42.644\\
add100&202&1100&28.469&0.112&254.582&8560&28562&&19.039&\\
add200&402&2200&118.100&0.194&608.789&17160&57262&&78.679&\\
add300&602&3300&267.400&0.274&976.925&25760&85962&&176.396&\\
add400&802&4400&481.833&0.442&1088.953&34360&114662&&314.843&\\
add500&1002&5500&750.889&0.582&1289.897&42960&143362&&502.329&\\
add600&1202&6600&1087.130&0.702&1549.569&51560&172062&&775.861&\\
add700&1402&7700&1492.829&0.837&1783.976&60160&200762&&1033.970&\\
add800&1602&8800&1945.277&0.998&1950.002&68760&229462&&1261.254&\\
add900&1802&9900&&1.017&&77360&258162&&1621.535&\\
add1000&2002&11000&&1.143&&85960&286862&&1927.034&
\end{tabular}
\caption{Evaluation results.  Quantum circuit adders consisting of \emph{Wires} wires and having \emph{Gates} gates are manipulated. The execution times (seconds) of the naive (\emph{NaiveT}) and the fast (\emph{FastT}) ICM transformations are compared. The transformations result in circuits consisting of \emph{ICM Wires} wires and \emph{ICM Ops} gates. The improvement columns (\emph{ImprT} and \emph{ImprR}) indicate the speed-up achieved by the fast manipulations.}
\label{tbl}
\end{table*}

\begin{figure}
    \centering
    \small{a)} \includegraphics[width=0.4\linewidth]{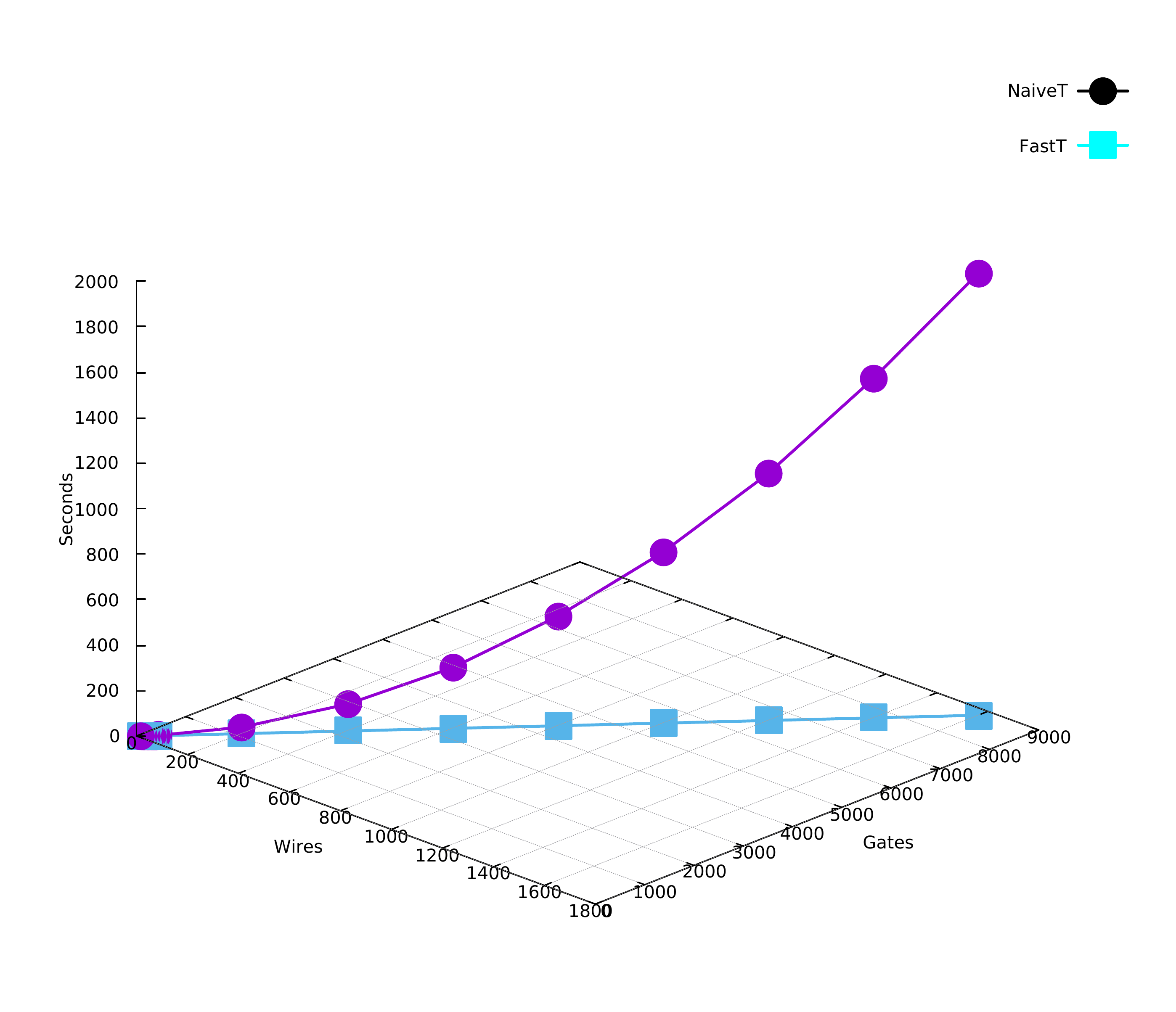}
    \small{b)} \includegraphics[width=0.4\linewidth]{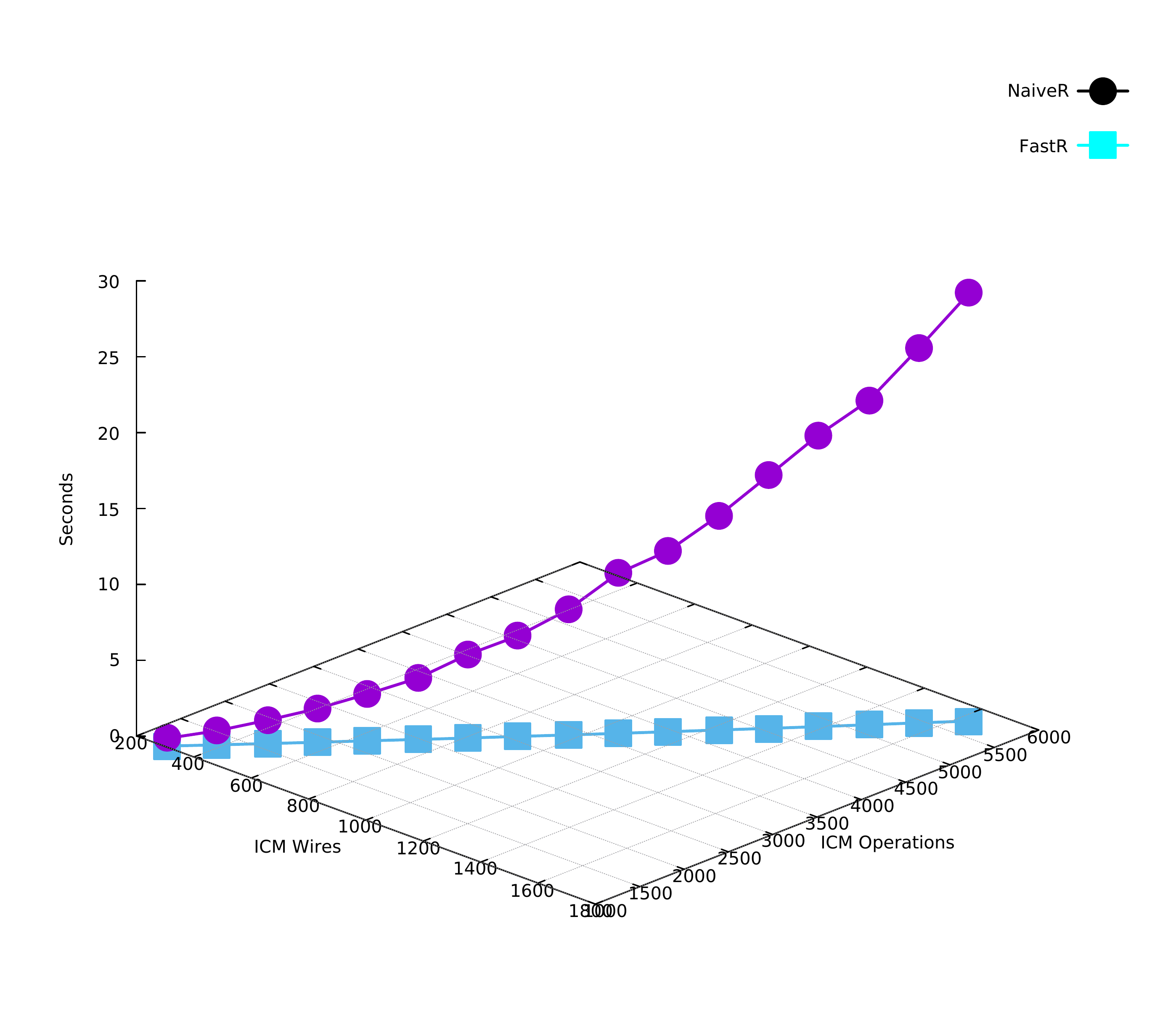}
    \caption{Plots of the results generated for the Table~\ref{tbl} rows where data was available (no missing entries). a) The \emph{Wires} and \emph{Gates} columns are on the X and Y axis, and the time needed to transform a Clifford+T circuit into ICM is illustrated on the Z axis. The data points for the naive and the fast transformation methods are plotted as lines in the 3D space. The speed-up column (\emph{ImprT} is not plotted.) b) The \emph{ICM Wires} and \emph{ICM Ops} columns are on the X and Y axis, and the time needed to transform a Clifford+T circuit into ICM is illustrated on the Z axis. The data points for the naive and the fast reordering/recycling methods are plotted as lines in the 3D space. The speed-up column (\emph{ImprR} is not plotted.)}
    \label{fig:plots}
\end{figure}

The performance achieved by using diagrams instead of naive wire label updates was investigated using quantum addition circuits consisting entirely of Toffoli gates \cite{cuccaro2004new}. Adders form the basis of arithmetic quantum algorithms, which are of particular practical interest. Simultaneously, the Toffoli gate is the preferred building block in the classical reversible circuit research community. A single hardware platform is known to be able to execute Toffoli based circuits \cite{chamon2017quantum}, but it can be assumed that their large scale execution will be performed on universal quantum computers. Furthermore, for the time being it is assumed that a large scale error corrected quantum computer will execute Clifford+$T$ computations, which have to be decomposed into a surface code compatible form (ICM). Therefore, the overhead of decomposing Toffoli gates, using e.g. rewrite and reorder operations, into Clifford+$T$ and ICM is benchmarked.

This evaluation approach contrasts to synthetic benchmarks using randomly generated circuits, or benchmarks consisting of circuits formed of generalised Toffoli gates (more than three inputs/outputs, e.g. RevKit). In order to determine the practicality of the data structure, the latter circuits would need to be first decomposed into three qubit Toffoli gates. Afterwards the evaluation would be similar to the one chosen. As a result, the evaluation is executed as follows: (1)~rewriting the Toffoli gate based adders into the equivalent ICM form \cite{paler2017fault}, and (2)~recycling the wires of the ICM circuits (reordered) \cite{paler2016wire}. Rewriting is used to obtain the fault-tolerant form of the adders, and recycling to optimise them without affecting their initialisations, gate lists or measurements. Thus, the circuits are still fault-tolerant.

The evaluation results are presented in Table~\ref{tbl} and Fig.~\ref{fig:plots}. The execution times of the naive recycling implementation (\emph{NaiveR}) and of the fast (\emph{FastR}) are compared. The improvement columns (\emph{ImprT} and \emph{ImprR}) indicate the speed-up achieved by the fast manipulations. Both the naive and the fast methods result in the same circuits, but the methods have different execution times (expressed in seconds).

The empiric evaluations were performed on an i5 machine with 8GB memory, and executions longer than 2000 seconds were timed out. The results indicate that the achieved speed-ups have the same magnitude with the number of wires of the circuit being manipulated. For example, the add800 circuit has 1602 wires before being rewritten in ICM, and the new rewrite method is 1950 times faster than the naive implementation.

It should be noted that reordering (recycling) benefited not only from the wire label diagrams, but also from the improved reachability analysis (see Sec.~\ref{sec:reach}). \emph{The naive implementation used standard set libraries for storing and updating the reached outputs, whereas binary arrays are much faster.}

The memory footprint is the performance bottleneck of the reachability analysis. It is possible to reduce the footprint by using data structures optimised for bit sets (e.g. \cite{lemire2016consistently}), but for this work it was considered that the current performance was sufficient. It was possible to automatically design circuits with 10 000 qubits. 

The results show that currently employed quantum circuit manipulation methods are computationally intensive although their complexity is only polynomial. There are sufficiently many possible improvements to achieve faster quantum circuit manipulation, and the herein presented wire label reference diagrams improve design automation performance.

\section{Discussion}

Wire label reference diagrams were motivated by wire labels being references and not order identifiers. As illustrated by the evaluation results and discussed in the previous example, this approach saves a lot of unnecessary updates. This achieves significant speed-ups for large fault-tolerant quantum circuit compilation and optimisation.

\begin{figure}
  \centering
  \includegraphics[width=0.7\columnwidth]{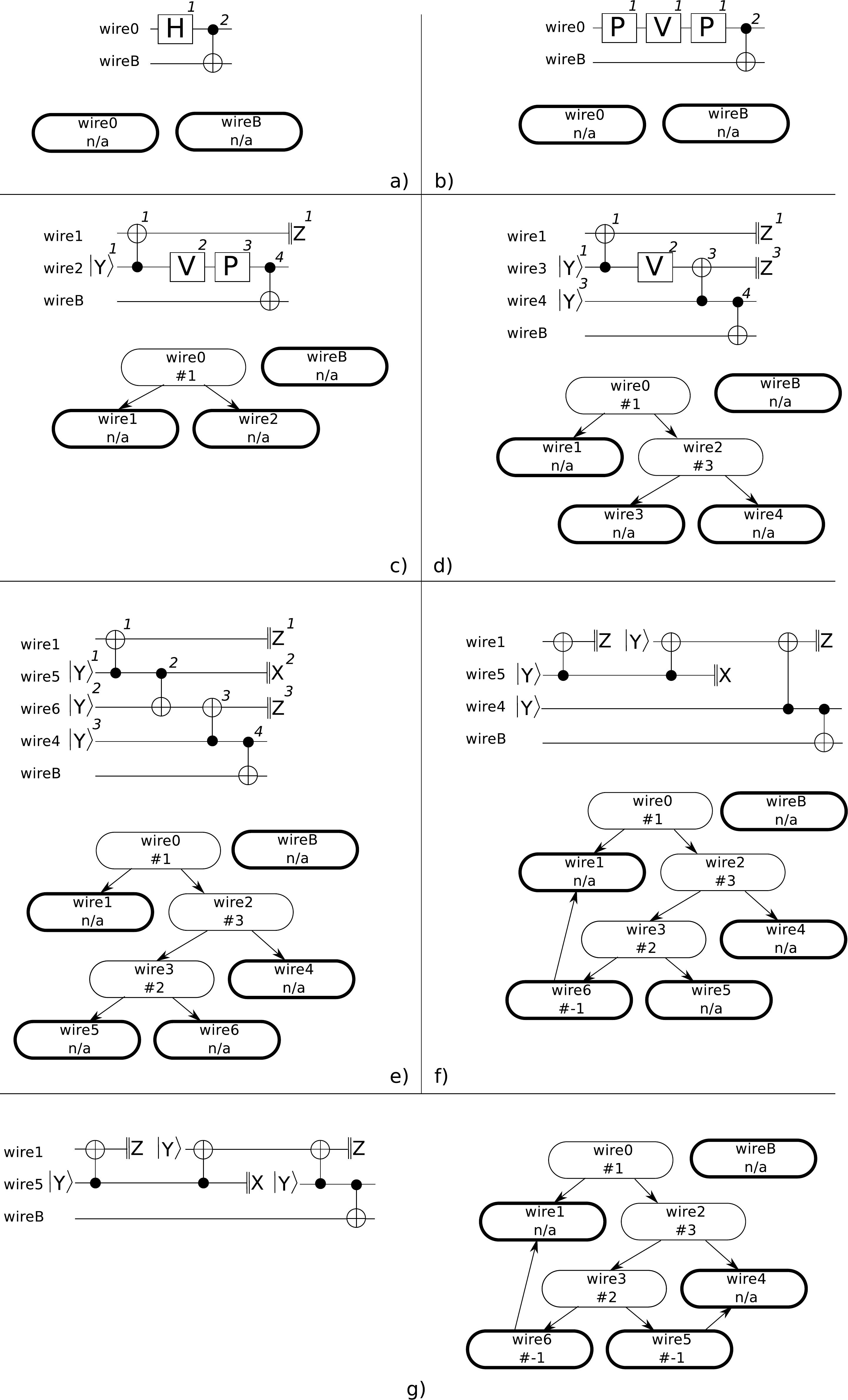}
  \caption{Example of using wire label reference diagrams: a) the initial circuit and diagram; b) - d) results after rewrite manipulations; e) - g) results after reordering manipulations. The most up-to-date wire label node has a thicker stroke. Note that in f) and g) the nodes wire5 and wire6 have swapped positions to simplify the resulting diagram.}
  \label{fig:disc}
\end{figure}

\subsection{Diagrammatic (global) vs. per circuit operation (local) view}

It is possible to argue against the diagrammatic representation of a circuit's structural changes. Instead of storing wire label changes into a diagram, one could directly update the references associated with the circuit operations. In this section we compare \emph{local} with \emph{global} perspectives of the circuit and use Fig.~\ref{fig:disc} to highlight the advantages of the proposed diagrams.

On the one hand, such an approach is local and straightforward according to the operation model from Section~\ref{sec:def} and the procedure from Fig.~\ref{fig:diag1}: if an operation uses a label reference equal to wire0, depending on its identifier (\#gatex), one of the new references (wire1 or wire2) is updated (written) in the reference list. No references are updated otherwise. This approach affects circuit operations individually without considering any other identical or similar circuit manipulations. The local approach has some advantages because of its simplicity: updates are immediate (eager), no additional data structure is needed (the reference diagram). Nevertheless, eager updates are time consuming because they do not take into consideration if: a) an update is necessary; b) memory writes are more expensive than memory reads.

On the other hand, the diagrams offer a global perspective about the circuit manipulation results, because each node and edge indicate how sets of circuit operations are affected. The simplest example is the diagram from Fig.~\ref{fig:disc}, where all operations with an identifier higher than the threshold of wire0 will use the reference wire2. Therefore, the diagram abstracts multiple updates by a single path. 

Diagrams have at least two advantages. First, the updates are lazy: they are performed only when necessary. Thus, multiple updates can be combined into a single one that is actually performed. Second, the diagrams offer the necessary support for \emph{update result reuse}: only a first operation from a specific operation needs to call an update, and the other operations can just reuse the update result. Reuse would employ a per node cache-similar mechanism implemented through dictionaries: the most relevant paths searched from that particular node are stored in the dictionary. Relevant could mean longest, most searched etc. However, being a cache, cache coherence methods need to be implemented. This work, and the results from Section~\ref{sec:res}, did not implement and consider such mechanisms, because of their complexity and empiric calibration requirements. However, the current results illustrate the speed-up potential of future label reference diagram improvements. As a result, the global and diagrammatic perspective of the circuit is better than the local one and is advantageous even when implemented straightforwardly.

\subsection{Analogies to classical circuit design automation}

Wire label reference diagrams share similarities to binary decision diagrams (BDD), which are a thoroughly investigated topic \cite{bryant1992symbolic} in computer science. The canonical interpretation of BDDs is related to Boolean functions and their normal forms. The state space of an n-bit classical circuit is $2^n$, and BDDs are used to compress losslessly the exponential representational overhead in order to make circuit manipulations more efficient. The compression is not often successful, but there are well documented practical scenarios of efficient manipulations of (conventional) function representations as well as circuits \cite{hachtel2006logic}.

Functional representations of quantum circuits are reliant on reducing the amount of representational redundancy. For example, quantum multiple-valued decision diagrams (QMDDs) \cite{miller2006qmdd} are decision diagrams (similar to binary decision diagrams) for representing the matrices associated to the quantum gates \cite{NC00}. 

Wire label reference diagrams are used for decisions too, but not from an exponentially large state space, because the search of the updated labels is a tree traversal using also if-then-else logic. Reference diagrams diagrams, in contrast to BDDs, do not compress the circuit's state representational overhead, but are a structured way of organising the results of quantum circuit manipulation. BDDs are used for functional decisions, whereas reference diagrams for structural decisions.

The reachability analysis uses the concept of cones, which are very similar to how it is employed in classical circuit design automation. For example, a \emph{fault cone} is the portion of a circuit whose signals are reachable by a forward trace of the circuit topology starting at the fault site \cite{bushnell2004essentials}. This work used a backward trace (from outputs to inputs), which implies that the analysis is not unsimilar to the single-output cones resulting from circuit segmentation \cite{bushnell2004essentials}.

\section{Conclusion}
\label{sec:concl}

The purpose of this work was to present the problem of large scale quantum circuit manipulation, and to highlight the potential of straightforward improvements. Future work will focus on a complexity analysis of the wire label reference diagrams, and on methods to augment them for even larger circuits.

There are multiple possible improvements regarding the performance of quantum circuit manipulation ranging from optimising the data structures for reachability analysis to reducing the number of algorithmic steps in the manipulation procedures. Future work will also consider fine tuning the methods.

\section{Acknowledgements}
AP acknowledges support through the Linz Institute of Technology project CHARON. This work has partially been supported by the European Union through the COST Action IC1405.

\bibliographystyle{unsrt}
\bibliography{telewrite} 

\begin{thebibliography}{10}

\bibitem{mohseni2017commercialize}
Masoud Mohseni, Peter Read, Hartmut Neven, Sergio Boixo, Vasil Denchev, Ryan
  Babbush, Austin Fowler, Vadim Smelyanskiy, and John Martinis.
\newblock {Commercialize quantum technologies in five years.}
\newblock {\em Nature}, 543(7644):171--174, 2017.

\bibitem{saeedi2013synthesis}
Mehdi Saeedi and Igor~L Markov.
\newblock {Synthesis and optimization of reversible circuits -- a survey}.
\newblock {\em ACM Computing Surveys (CSUR)}, 45(2):21, 2013.

\bibitem{shende2003synthesis}
Vivek~V Shende, Aditya~K Prasad, Igor~L Markov, and John~P Hayes.
\newblock {Synthesis of reversible logic circuits}.
\newblock {\em Computer-Aided Design of Integrated Circuits and Systems, IEEE
  Transactions on}, 22(6):710--722, 2003.

\bibitem{DBLP:conf/aspdac/NiemannWD14}
Philipp Niemann, Robert Wille, and Rolf Drechsler.
\newblock {Efficient synthesis of quantum circuits implementing Clifford group
  operations}.
\newblock In {\em Asia and South Pacific Design Automation Conference}, pages
  483--488, 2014.

\bibitem{miller2003transformation}
D~Michael Miller, Dmitri Maslov, and Gerhard~W Dueck.
\newblock {A transformation based algorithm for reversible logic synthesis}.
\newblock In {\em Design Automation Conference}, pages 318--323, 2003.

\bibitem{DBLP:conf/ismvl/MillerWD10}
D.~Michael Miller, Robert Wille, and Rolf Drechsler.
\newblock {Reducing Reversible Circuit Cost by Adding Lines}.
\newblock In {\em International Symposium on Multiple-Valued Logic}, pages
  217--222, 2010.

\bibitem{DBLP:journals/integration/WilleSMD14}
Robert Wille, Mathias Soeken, D.~Michael Miller, and Rolf Drechsler.
\newblock {Trading off circuit lines and gate costs in the synthesis of
  reversible logic}.
\newblock {\em Integration}, 47(2):284--294, 2014.

\bibitem{paler2016wire}
Alexandru Paler, Robert Wille, and Simon~J Devitt.
\newblock {Wire recycling for quantum circuit optimization}.
\newblock {\em Physical Review A}, 94(4):042337, 2016.

\bibitem{hirata2011efficient}
Yuichi Hirata, Masaki Nakanishi, Shigeru Yamashita, and Yasuhiko Nakashima.
\newblock {An efficient conversion of quantum circuits to a linear nearest
  neighbor architecture}.
\newblock {\em Quantum Information \& Computation}, 11(1\&2):142--166, 2011.

\bibitem{DBLP:journals/tcad/WilleLD14}
Robert Wille, Aaron Lye, and Rolf Drechsler.
\newblock {Exact Reordering of Circuit Lines for Nearest Neighbor Quantum
  Architectures}.
\newblock {\em {IEEE} Trans. on {CAD} of Integrated Circuits and Systems},
  33(12):1818--1831, 2014.

\bibitem{barenco1995elementary}
Adriano Barenco, Charles~H Bennett, Richard Cleve, David~P DiVincenzo, Norman
  Margolus, Peter Shor, Tycho Sleator, John~A Smolin, and Harald Weinfurter.
\newblock {Elementary gates for quantum computation}.
\newblock {\em Physical review A}, 52(5):3457, 1995.

\bibitem{NC00}
Michael~A Nielsen and Isaac~L Chuang.
\newblock {\em {Quantum computation and quantum information}}.
\newblock Cambridge university press, 2010.

\bibitem{DBLP:conf/rc/NiemannBCJW15}
Philipp Niemann, Saikat Basu, Amlan Chakrabarti, Niraj~K. Jha, and Robert
  Wille.
\newblock {Synthesis of Quantum Circuits for Dedicated Physical Machine
  Descriptions}.
\newblock In {\em International Conference Reversible Computation}, pages
  248--264, 2015.

\bibitem{paler2017fault}
Alexandru Paler, Ilia Polian, Kae Nemoto, and Simon~J Devitt.
\newblock {Fault-tolerant, high-level quantum circuits: form, compilation and
  description}.
\newblock {\em Quantum Science and Technology}, 2(2):025003, 2017.

\bibitem{FMM13}
A.G. Fowler, M.~Mariantoni, J.M. Martinis, and A.N. Cleland.
\newblock {Surface Codes, Towards practical large-scale quantum computation}.
\newblock {\em Phys. Rev. A.}, 86:032324, 2012.

\bibitem{soeken2012revkit}
Mathias Soeken, Stefan Frehse, Robert Wille, and Rolf Drechsler.
\newblock {RevKit: A Toolkit for Reversible Circuit Design.}
\newblock {\em Multiple-Valued Logic and Soft Computing}, 18(1):55--65, 2012.

\bibitem{anders2006fast}
Simon Anders and Hans~J Briegel.
\newblock {Fast simulation of stabilizer circuits using a graph-state
  representation}.
\newblock {\em Physical Review A}, 73(2):022334, 2006.

\bibitem{balensiefer2005evaluation}
Steven Balensiefer, Lucas Kregor-Stickles, and Mark Oskin.
\newblock {An evaluation framework and instruction set architecture for
  ion-trap based quantum micro-architectures}.
\newblock In {\em ACM SIGARCH Computer Architecture News}, volume~33, pages
  186--196. IEEE Computer Society, 2005.

\bibitem{maslov2008quantum}
Dmitri Maslov, Gerhard~W Dueck, D~Michael Miller, and Camille Negrevergne.
\newblock {Quantum circuit simplification and level compaction}.
\newblock {\em IEEE Transactions on Computer-Aided Design of Integrated
  Circuits and Systems}, 27(3):436--444, 2008.

\bibitem{zulehner2017exact}
Alwin Zulehner, Stefan Gasser, and Robert Wille.
\newblock {Exact Global Reordering for Nearest Neighbor Quantum Circuits Using
  A*}.
\newblock In {\em International Conference on Reversible Computation}, pages
  185--201, 2017.

\bibitem{DBLP:conf/date/ZulehnerW17}
Alwin Zulehner and Robert Wille.
\newblock {Make it reversible: Efficient embedding of non-reversible
  functions}.
\newblock In {\em Design, Automation {\&} Test in Europe Conference}, pages
  458--463, 2017.

\bibitem{van2016path}
Rodney Van~Meter and Simon~J Devitt.
\newblock {The path to scalable distributed quantum computing}.
\newblock {\em Computer}, 49(9):31--42, 2016.

\bibitem{fowler2012time}
Austin~G Fowler.
\newblock {Time-optimal quantum computation}.
\newblock {\em arXiv preprint arXiv:1210.4626}, 2012.

\bibitem{bishop2017quantum}
Lev~S Bishop, Sergey Bravyi, Andrew Cross, Jay~M Gambetta, and John Smolin.
\newblock {Quantum Volume}.
\newblock
  https://dal.objectstorage.open.softlayer.com/v1/AUTH\_039c3bf6e6e54d76b8e66152e2f87877/community-documents/quatnum-volumehp08co1vbo0cc8fr.pdf,
  2017.

\bibitem{ross2014optimal}
Neil~J Ross and Peter Selinger.
\newblock {Optimal ancilla-free Clifford+T approximation of z-rotations}.
\newblock {\em arXiv preprint arXiv:1403.2975}, 2014.

\bibitem{selinger2013quantum}
Peter Selinger.
\newblock {Quantum circuits of T-depth one}.
\newblock {\em Physical Review A}, 87(4):042302, 2013.

\bibitem{cuccaro2004new}
Steven~A Cuccaro, Thomas~G Draper, Samuel~A Kutin, and David~Petrie Moulton.
\newblock {A new quantum ripple-carry addition circuit}.
\newblock {\em arXiv preprint quant-ph/0410184}, 2004.

\bibitem{chamon2017quantum}
Claudio Chamon, ER~Mucciolo, AE~Ruckenstein, and Z-C Yang.
\newblock {Quantum vertex model for reversible classical computing}.
\newblock {\em Nature Communications}, 8, 2017.

\bibitem{lemire2016consistently}
Daniel Lemire, Gregory Ssi-Yan-Kai, and Owen Kaser.
\newblock {Consistently faster and smaller compressed bitmaps with roaring}.
\newblock {\em Software: Practice and Experience}, 46(11):1547--1569, 2016.

\bibitem{bryant1992symbolic}
Randal~E Bryant.
\newblock {Symbolic Boolean manipulation with ordered binary-decision
  diagrams}.
\newblock {\em ACM Computing Surveys (CSUR)}, 24(3):293--318, 1992.

\bibitem{hachtel2006logic}
Gary~D Hachtel and Fabio Somenzi.
\newblock {\em {Logic synthesis and verification algorithms}}.
\newblock Springer Science \& Business Media, 2006.

\bibitem{miller2006qmdd}
D~Michael Miller and Mitchell~A Thornton.
\newblock {QMDD: A decision diagram structure for reversible and quantum
  circuits}.
\newblock In {\em Multiple-Valued Logic, 2006. ISMVL 2006. 36th International
  Symposium on}, pages 30--30. IEEE, 2006.

\bibitem{bushnell2004essentials}
Michael Bushnell and Vishwani Agrawal.
\newblock {\em {Essentials of electronic testing for digital, memory and
  mixed-signal VLSI circuits}}, volume~17.
\newblock Springer Science \& Business Media, 2004.

\end{thebibliography}

\end{document}